\documentclass[useAMS,usenatbib]{mn2e}

\usepackage{graphicx,amssymb,bm}
\usepackage{subfig,xfrac}
\usepackage{float, Abbrev}
\usepackage[fleqn]{amsmath}  
\usepackage{color}

\usepackage{multirow}

\newcommand{\be}{\begin{equation}}
\newcommand{\ee}{\end{equation}}
\newcommand{\ba}{\begin{eqnarray}}
\newcommand{\ea}{\end{eqnarray}}

\newcommand{\lcdm}{$\Lambda$CDM }

\def\simlt{\lower.5ex\hbox{$\; \buildrel < \over \sim \;$}}

\newcommand{\fig}{\begin{figure} \begin{center}}
\newcommand{\efig}{\end{center}\end{figure} }
\newcommand{\figs}{\begin{figure*}\begin{minipage}{180mm} \begin{center}}
\newcommand{\efigs}{\end{center}\end{minipage}\end{figure*} }
\def\simgt{\lower.5ex\hbox{$\; \buildrel > \over \sim \;$}}

\usepackage{graphicx} 

\title[Dark matter misalignments in HFF]{Systematic or Signal? How dark matter misalignments can bias strong lensing models of galaxy clusters}
\author[D. Harvey, J. P. Kneib \& M. Jauzac]
{D. Harvey$^{1}$\thanks{e-mail: {\tt david.harvey@epfl.ch}}, J. P. Kneib$^{1,2}$ and M. Jauzac$^{3,4,5}$ \\
$^{1}$Laboratoire d'Astrophysique, Ecole Polytechnique F$\acute{e}$d$\acute{e}$rale de Lausanne (EPFL), Observatoire de Sauverny, CH-1290 Versoix, Switzerland \\
$^{2}$Aix Marseille Université, CNRS, LAM (Laboratoire d'Astrophysique de Marseille) UMR 7326, 13388, Marseille, France \\
$^{3}$Centre for Extragalactic Astronomy, Department of Physics, Durham University, Durham DH1 3LE, U.K.\\
$^{4}$Institute for Computational Cosmology, Durham University, South Road, Durham DH1 3LE, U.K. \\
$^{5}$Astrophysics and Cosmology Research Unit, School of Mathematical Sciences, University of KwaZulu-Natal, Durban 4041, South Africa}
\begin{document}

\date{Accepted ---. Received ---; in original form \today.}

\pagerange{\pageref{firstpage}--\pageref{lastpage}} \pubyear{2013}

\maketitle

\label{firstpage}

\begin{abstract}
\noindent 
We explore how assuming that mass traces light in strong gravitational lensing models can lead to systematic errors in the predicted position of multiple images. 
Using a model based on the galaxy cluster MACSJ0416 ($z=0.397$) from the Hubble Frontier Fields, we split each galactic halo into a baryonic and dark matter component.
We then shift the dark matter halo such that it no longer aligns with the baryonic halo and investigate how this affects the resulting position of multiple images.
We find for physically motivated misalignments in dark halo position, ellipticity, position angle and density profile, that multiple images can move on average by more than $0.2\arcsec$ with individual images moving greater than $1\arcsec$. 
 We finally estimate the full error induced by assuming that light traces mass and find that this assumption leads to an expected RMS error of $0.5\arcsec$, almost the entire error budget observed in the Frontier Fields.
Given the large potential contribution from the assumption that light traces mass to the error budget in mass reconstructions, we predict that it should be possible to make a first significant detection and characterisation of dark halo misalignments in the Hubble Frontier Fields with strong lensing.
Finally, we find that it may be possible to detect $\sim1$kpc offsets between dark matter and baryons, the smoking gun for self-interacting dark matter,  should the correct alignment of multiple images be observed.

\end{abstract}

\begin{keywords}
cosmology: dark matter --- galaxies: clusters --- gravitational lensing
\end{keywords}

\section{Introduction}
Mapping the distribution of total matter in galaxy clusters has become commonplace with the advent of high resolution optical imaging from space \cite[e.g.][]{A2744,MACSJ0717_HFF,A2744_HFF}.
Deep images of galaxy clusters reveal the apparent distortion of distant background galaxies whose light has been split into many geodesics producing multiple images of the same galaxy. 
Strong gravitational lensing has become a vital tool in mapping out the distribution of  matter in galaxy clusters as well as its behaviour during highly energetic collisions \cite[e.g.][]{bulletclusterB,A2744}. For a review see \cite{gravitational_lensing}.

Methods to reproduce the distribution of matter in galaxy clusters can be split into two categories: those that assume that light traces mass \citep[e.g.][]{zitrin,lenstool} and those that do not \citep[e.g.][]{bradac,grale,merten_SL}. Under the light traces masses assumption it is assumed that wherever there is a galaxy there also exists a dark matter halo, the mass of which far out exceeds that of the baryonic component \cite[e.g.][]{A1689}. Specifically, the assumption is that the peak of the dark matter halo lies exactly coincident with the galaxy, with an equal ellipticity and equal position angle, the only difference in being the scale at which the light and dark halos extend to. The main advantage of assuming that mass traces light is that by scaling the DM halo directly to the light distribution of the galaxy it is possible to significantly reduce the number of free parameters in a strong lensing model of a galaxy cluster that may contain up to a few hundred individual galaxies. This reduces the computational time for a single reconstruction and increases the constraining power of the model by placing heavy priors on each galactic halo. However, conversely such an assumption may lead to inaccuracies whereas free-form reconstructions that do not assume this may not impose a bias but will suffer in precision. It is in this paper that we explicitly examine these underlying assumptions. We ask whether these assumptions are suitable and in the event that they are not, how do they affect the small-scale properties of strong lensing models used. 
Similarly, we also determine whether the misalignment of dark matter could be detected for the first time with current data.

\fig
\includegraphics[width=0.45\textwidth]{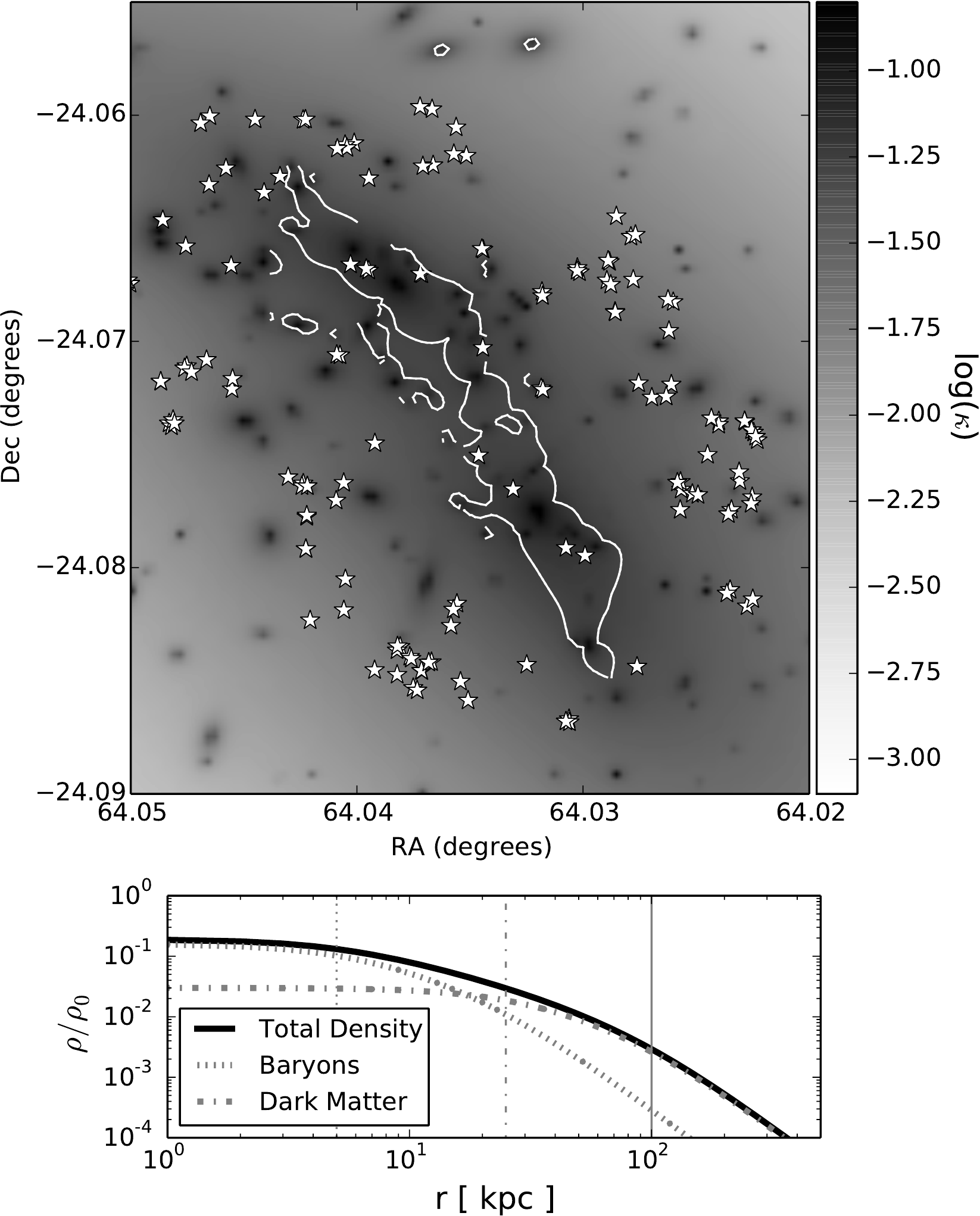} 
\caption{\label{fig:model} The cluster model we used. The top panel shows the galaxy cluster with the convergence map in grey scale and the resulting critical lines in white. The white stars represent the true position of multiple images. The bottom panel shows an example of how we split a galaxy with a total PIEMD profile of $r_{\rm core}=5$ kpc and $r_{\rm cut}=100$ kpc, into 2 PIEMDs of $r^{\rm B}_{\rm core}=5$ kpc (vertical dotted line) , $r^{\rm B}_{\rm cut}=25$ kpc and $r^{\rm DM}_{\rm core}=25$ kpc (vertical dot-dashed line), $r^{\rm DM}_{\rm cut}=100$ kpc (vertical solid line) .}
\efig

\subsection{Does light trace mass?}

In order to motivate the study of how assuming mass traces light affects strong lensing models, we must first question whether we would expect the profile of dark matter halo to mimic that of the baryonic component. In a \lcdm Universe, cosmological simulations predict that dark matter should lie coincident with the baryonic component, with no physical offset of $\delta r>3$kpc  being observed \citep{LCDM_offsets}. However, extensions to the collisionless cold dark matter model have been proposed that could lead to offsets between baryonic and dark component \citep{SIDMModel,Harvey14}. For example, \cite{cannibal} and subsequently \cite{A3827_massey} observed a $1.5$kpc offset between the two components in an elliptical galaxy in the cluster Abell 3827, at the $3\sigma$ level.  This offset was attributed to potential dark matter self-interactions causing a lag on the dark matter halo. It is therefore possible that dark matter could indeed separate from its baryonic counterpart. 

Aside from the peak position of the dark matter halo, it isn't clear whether the dark matter halo should mimic the geometrical properties of the baryonic component.
A recent study using the MassiveBlack II cosmological hydrodynamical simulations found that this was not true \citep{bary_dm_allign}. 
They found that dark matter halos tended to be rounder, by up to factors of two, and that the major axes could become severely misaligned with offsets of $\theta=90^\circ$ not uncommon. They found that the offsets were most common at galaxy scale masses, with larger, cluster size halos better aligned. These findings are consistent with others in the field \citep[e.g.][]{bary_dm_allign2,bary_dm_allign3,bary_dm_allign_owls}. 
Additionally, assuming that light traces mass often assumes an empirical relation between the size of the dark matter halo and the luminosity \cite[e.g.][]{A1689}.
Although well constrained, these relations can exhibit a variance of up to $50\%$ \citep{mass_luminosity,mass_luminosity2}.
Given that simulations predict that dark matter and baryonic halos should not necessarily trace each other we are motivated to test how this assumption can affect the strong lensing models. 

Previous work studying the alignment of dark with light matter using strong lensing is limited. \cite{align_SL} first studied how multiple image separations can be altered by misalignments between the dark matter and galactic halo. They showed how the statistics of image separation can avoid the need to model each lens and hence can be useful for large scale surveys. More recently \cite{align_SL_2} carried out a study of 11 lensing galaxies examining their alignment with their baryonic component. Similar to simulations they found that halos are rounder than their galactic counterpart and those galaxies with large amounts of shear are highly misaligned. Although very interesting, this is limited to small number of galaxies in groups. In this work we will look at carrying out a similar study except over large numbers of galaxies that reside within the high density cluster environment.
\section{Method}

\figs
       \includegraphics[width=\textwidth]{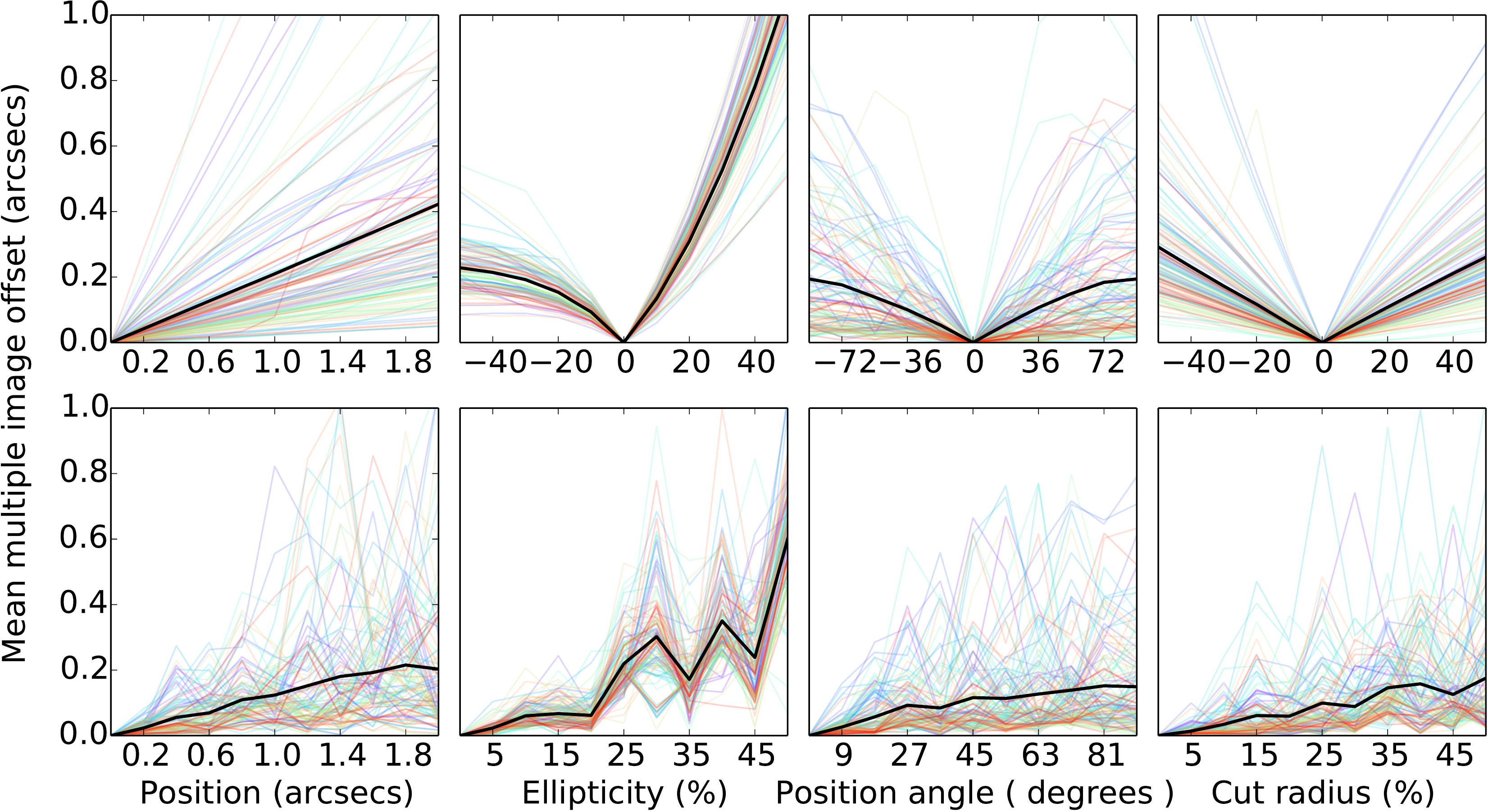} 
       	\caption{  \label{fig:profiles} Results from MACSJ0416 simulation. Each case the faint lines show the track of individual images, and the black line represents the RMS of all of the images with respect to the original catalogue. In each case in the top row the dark matter component in all the galaxies have been systematically changed (such that all galaxies have the same offset), and the bottom row gives the standard deviation of the normal distribution from which the offset has been randomly selected (such that each galaxy in the model has a different offset). In the case of the position offset (first column and top row) each galaxy has had their dark matter halo pushed away from the centre of mass of the cluster.}
	\efigs

To study the effect that assuming light traces mass has we use a model of a cluster based on real data that contains 174 small scale galaxy-scale halos \citep[spectroscopically identified by][]{MACSJ0416_spec} embedded in two large, cluster-scale dark matter halos \cite{MACSJ0416_HFF,MACSJ0416_spec}. 
In this study we act only to study the effect of changes to the galaxy scale halos on the positions of multiple images.
We use the position of multiple images in the Hubble Frontier Field galaxy cluster, MACSJ0416 and the best fitting cluster model that fits these multiple images as derived by the parametric strong lensing algorithm {\sc Lenstool} \citep{MACSJ0416_HFF,lenstool}. 
This equates to 140 multiple images, and 174 cluster members.
Each potential was originally fitted with a pseudo isothermal elliptical mass distribution (PIEMD) \citep{PIEMD,PIEMD2,PIEMD3,lenstool}, which follows the analytical density profile,
\be
\frac{\rho(r)}{\rho_0} = \frac{1}{\sqrt{r^2+r_{\rm core}^2}} - \frac{1}{\sqrt{r^2+r_{\rm cut}^2}},
\label{eqn:piemd}
\ee
where $r$ is the radial distance from the centre of the halo, and the profile is parameterised by the core radius, $r_{\rm core}$, and the cut radius, $r_{\rm cut}$.
The cluster itself has a measured mass within a $200$kpc aperture of $(1.6\pm0.01)\times10^{14}M_\odot$ \citep{MACSJ0416_HFF}, and the mass of each galaxy has a lognormal distribution centred around log$(M/M_\odot) = 10.1 \pm 0.6$. The distribution of ellipticities peaks at $0.1$ and decreases towards larger ellipiticites. The galaxies span the entire range between $0$ and $0.9$.
We split each galaxy halo potential into two components: a baryonic core and a dark matter halo. 
We separate the halo into two PIEMDs with the baryonic $r^{\rm B}_{\rm core} = r_{\rm core}$. 
For the baryonic cut radius, we conservatively cut it to a quarter of the original cut radius $r^{\rm B}_{\rm cut} = \sfrac{r_{\rm cut}}{4}$ \citep{dm_baryon_halo}, although this could be smaller. We also force the dark matter core radius to the baryonic cut radius, $r^{\rm DM}_{\rm core} = r^{\rm B}_{\rm cut}$ and the dark matter cut radius as the original cut radius $r^{\rm DM}_{\rm cut} = r_{\rm cut}$.
Separating the halo like this means that the mass of each halo is conserved with respect to the pre-split halo.
Given this two component galaxy model we project the images back to the source plane to create a list of background sources for the cluster.  
To do this we use the same redshift information for the sources as that in \cite{MACSJ0416_HFF}.

Figure \ref{fig:model} shows the makeup of the model. The top panel shows the cluster model with the position of the multiple images as white stars, log of the normalised projected surface density (convergence) map in grey scale and the critical lines in the white solid line. 
The bottom panel of Figure \ref{fig:model} gives the two PIEMD component model of the galaxy. 
The solid black line is a galaxy with a $r_{\rm core}=5$kpc and $r_{\rm cut}=100$kpc, and the resulting baryonic (dotted) and dark matter (dot-dashed) components when the galaxy is split in to two. 
The dotted, dot-dashed and solid vertical lines give the scales of the baryonic core radius, the dark matter core and baryonic cut radius and finally the dark matter cut radius respectively. 
\figs
       \includegraphics[width=\textwidth]{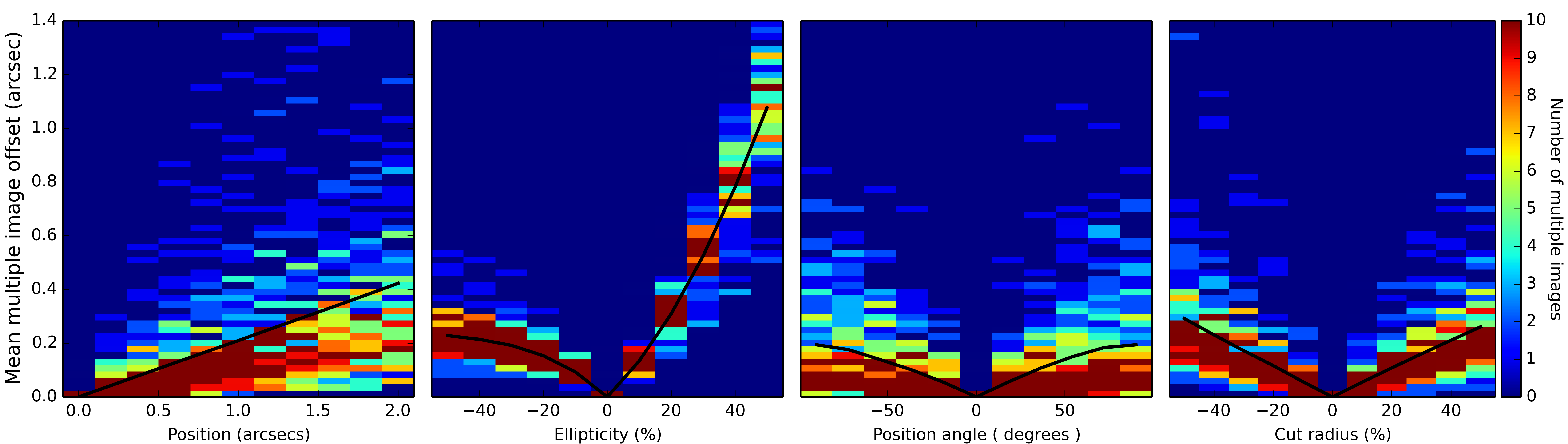}
       \caption{\label{fig:RMS} Histograms of the distribution of  multiple images w.r.t the RMS of all the images. In each case the systematic offset (top row of Figure \ref{fig:profiles}) is shown with the distribution of multiple images offset with the solid line showing the RMS of the multiple images. It can be clearly seen that the RMS is biased high by one or two individual spuriously offset halos.}
\efigs

To test the validity of assuming that light traces mass we shift the dark matter component of the lensing model however keep the baryonic component constant. We project the sources back through to the image plane with the new dark matter model and calculate the shift in position of each multiple image. We test four shifts in the dark matter component: 
\begin{enumerate}
\item Position: offsetting the peak position of the dark matter halo from the baryonic;
\item Ellipticity: stretching or squashing the dark matter halo with respect to the baryonic component;
\item Position angle: rotating the halo such that the dark matter and baryonic component are misaligned;
\item Cut radius: we vary the cut radius and velocity dispersion  of the dark matter halo to change the profile of the galaxy however conserve the total mass of the dark matter halo.
\end{enumerate}
Additionally, for each case we will either shift the halo systematically in the same direction, or randomly with a mean of zero and a standard deviation of the given value.

\section{Results}

We first test each assumption individually, beginning with peak position offset.
We start by systematically shifting the dark matter halos position away from the peak of the light distribution.
We shift the halo in a direction anti-parallel to the centre of mass of cluster, to imitate a shift in the dark matter halo due to self-interactions \citep[see][]{SIDMModel}.
The first panel on the top row of Figure \ref{fig:profiles} shows the mean multiple image offset as a function of dark matter halo offset.
In each case the solid black line gives the root mean square (RMS) of all the multiple images for a given offset, and the fainter coloured lines, the offset of each individual image.
We find that for a given offset $\delta p$, the mean multiple image offset, $\delta r_{\rm im} \approx 0.2\delta p$.

Following this, we then shift the position of the dark matter halo by a random amount, sampled from a normal distribution with a mean of zero and an increasing standard deviation. The first panel of the bottom row of Figure \ref{fig:profiles} gives the multiple image offset as a function of the input standard deviation.
We find that for random offsets with a standard deviation of $\delta p$, the mean multiple image offset is slightly less sensitive at, $\delta r_{\rm im} \approx 0.1\delta p$.

The following three tests study the sensitivity of multiple image position to a scalar quantity, and therefore unlike the position, do not require a reference point in which to be systematically offset.
We therefore show in the top row of Figure \ref{fig:profiles} how a common offset in ellipticity, position angle and mass can alter the position of multiple images (second, third and fourth column respectively) and the bottom row how a random offset, selected from a normal distribution with a mean of zero and an incrementally larger standard deviation can alter the mean position of multiple images.
 
Our tests reveal that highly elliptical dark matter halos result in very large shifts in the position of multiple images, with shifts of greater than one arc-second not uncommon.
More circular halos, also gave an offset, except less significant.
We also find that the changing position angle of the dark matter halo has sinusoidal relation, with some multiple images experiencing highly sensitive angles. 
Finally varying the cut radius but conserving mass has a symmetric effect, whether we decrease the cut radius and increase the velocity dispersion or vice versa.
We find that a shift in the position angle or cut radius result in similar magnitude offsets, both in systematically and randomly, with mean offsets of roughly $\sim0.2\arcsec$.

\subsection{Sensitivity of the model RMS to individual images}
The bottom row in Figure \ref{fig:profiles} shows that in all four tests carried out we find that the position of individual images can be very sensitive to misalignment of dark matter and baryons. 
This can be important since in a strong lensing reconstruction individual images can significantly bias the RMS value of predicted image position to actual image position. 
Not only this, but it may bias the rest of the reconstruction and any derived deflection maps. 
We illustrate this point further by binning the shift in multiple image position into a histogram and overlaying the RMS. 
Figure \ref{fig:RMS} shows the result for each test in the systematic offset mode (same as top row of Figure \ref{fig:profiles}). 
We see that in all cases except ellipticity, the thick solid line which represents the RMS is significantly above the majority of the actual image offsets. 
Particularly the offset due to position angle, whereby the majority are $<0.1\arcsec$ however some images at $0.8\arcsec$ can bias the RMS very high. 
However, interestingly, the RMS statistic well represents the error imposed by a an ellipticity offset.

\subsection{Sensitivity of image position to individual cluster members}

In order to better understand the origin of the RMS for each offset,  we take a multiple image that is particularly sensitive to a shift and we study its environment and how it changes with respect to the change in its dark matter halo. 
The top panel of Figure \ref{fig:image_4_2} ($20\arcsec\times20\arcsec$) shows how the position of one multiple image is sensitive to the position angle of nearby galaxies. 
We carry out a study whereby we rotate the nearby dark matter halos by incremental amounts, as represented by each coloured ellipse. 
For each incremental rotation we calculate the resulting position of the image, which we represent as a star whose colour matches that of the dark matter position angle that caused it.
In the bottom panel, the stars correspond to the stars of the same colour in the top panel except shown as a magnitude offset from the original position. 
The coloured diamonds shows the offset when {\it just the closest lens} is altered (as shown by the lens with the arrow).
It can be seen that when the major axis of the rotated dark matter halo aligns radially (points towards) with the image the resulting error in the image is largest, however when the galaxy lies tangentially then the error is very small.
This, also explains the smooth sinusoidal function observed in Figure \ref{fig:profiles}.
Furthermore, we find when we alter just a single galaxy (diamonds), this can contribute the majority of the error in the position of a multiple image, showing that this perturbative effect is a local one concerning only very near massive halos. 
We confirm this by testing the correlation between the observed offset of an image and distance to the closest lens weighted by its mass. We find a positive correlation for all offsets except the cut radius, for which we find no such dependence.

We finally test the three remaining shifts (ellipticity, position angle and cut radius) and how perturbing the closest lens affects the position of the image.
The inset in Figure \ref{fig:image_4_2} ($2\arcsec\times2\arcsec$) shows the results from these tests, with the arrow indicating the direction in which we shift the dark matter peak position. 
We find that when the ellipticity, position angle and dark matter peak position of the nearest lens is altered, the effect is a tangential movement on the image and completely degenerate. 
However, when we change the cut radius of the galaxy, this results in a radial movement of the multiple image. 
This orthogonal movement will allow future models to constrain which effect is causing the RMS offset.

Given that the positional test is only for a shift in one direction we test all other possible directions. Figure \ref{fig:image_4_2_pos} (same dimensions as Figure \ref{fig:image_4_2}) shows the behaviour of the multiple image when shifting the lens in every possible way. The coloured arrows define the offset direction of the potential and the corresponding coloured lines in the inset show the movement of the multiple image from the original position given by the star. 
We find that the movement of the image is always tangential with some small range of angle. We confirm this by studying other cluster members and find the same effect.   
\fig
       \includegraphics[width=0.45\textwidth]{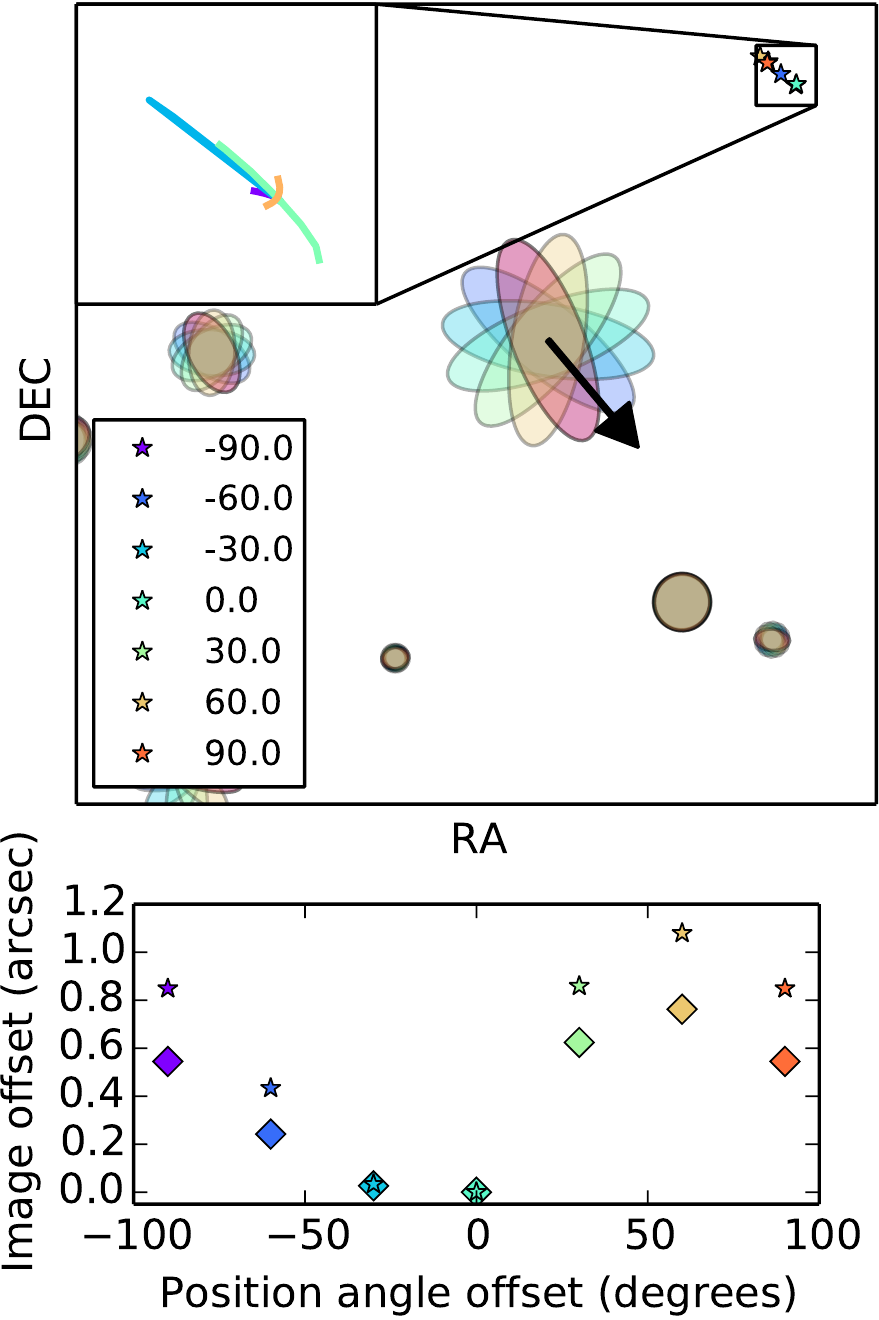} 
       \caption{ \label{fig:image_4_2} {\it Top :} Dependency of a given multiple image on the position angle of cluster member position angles. The main image ($20\arcsec\times20\arcsec$) gives the position of one multiple image as a function of cluster member position angle. The colours relate the position angle of the dark matter to the position of a multiple image, and how this corresponds to a radial offset from the original position in the bottom panel. The stars in this panel correspond to the star positions in the top, the diamonds give the image offset in the event that only the lens near the image is rotated. {\it Top panel Inset} ($2\arcsec\times2\arcsec$) shows how the same multiple image moves w.r.t different changes in the dark matter halo of only the lens nearest image.  Orange track shows the effect of changing the cut radius, green shows the movement due to changing the ellipticity, blue the position angle and purple the peak position of the dark matter halo.}
       
\efig	
     \fig
       \includegraphics[width=0.44\textwidth]{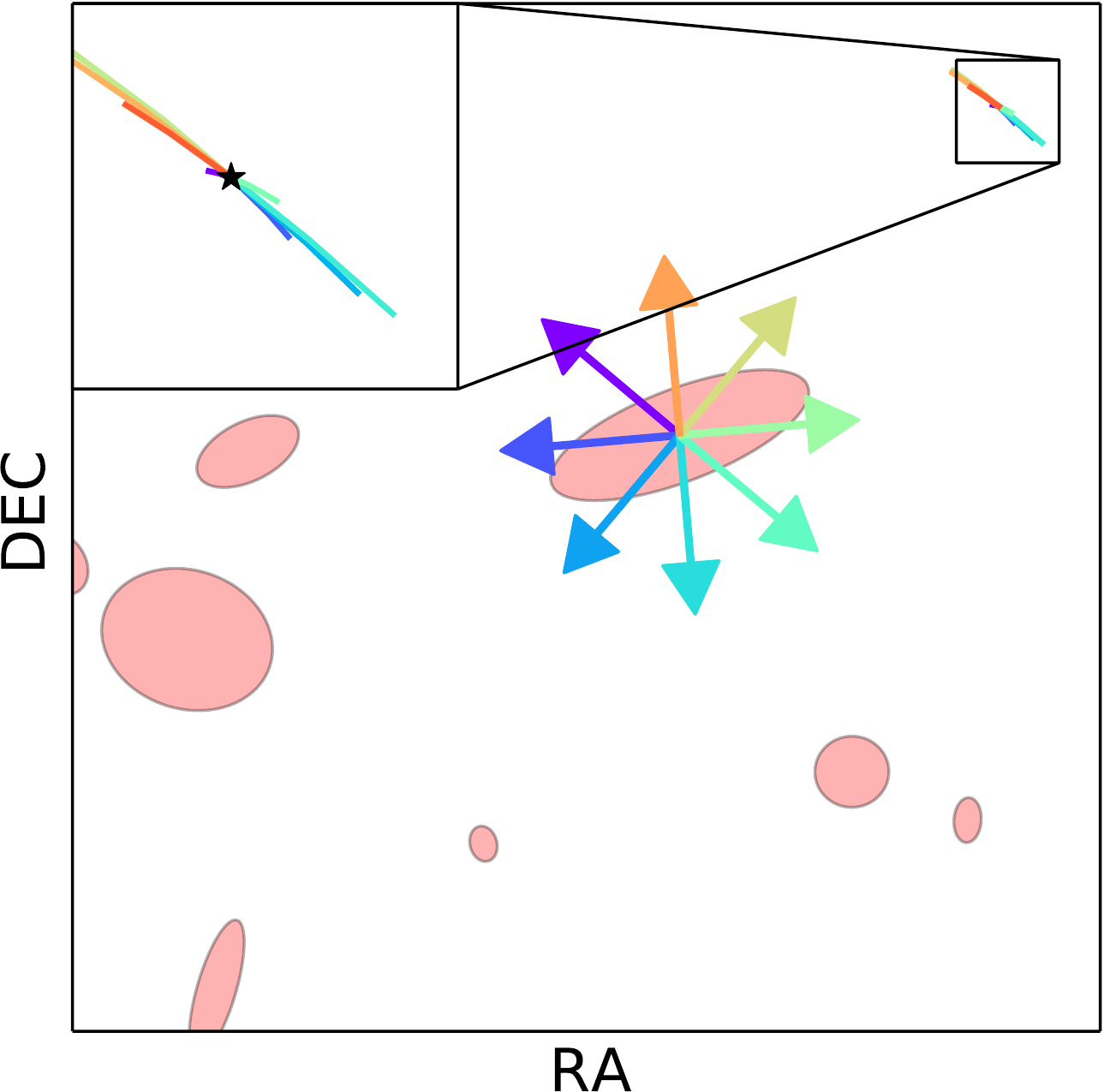}
       \caption{ \label{fig:image_4_2_pos} The same as Figure \ref{fig:image_4_2} (with the same dimensions) except this shows how shifting the dark matter halo peak position in any direction results only in a tangential movement of the multiple image.}
\efig	

\subsection{ Combined all the offsets: testing the assumption }
In practice all four assumptions will affect the predictability of a strong lensing model. Here we estimate how these offsets combine and what the expected RMS would be. To do this we produce many realisations of the same cluster using fixed offset parameters for the galaxy-scale halos and continuing to keep the large scale cluster halo fixed. We select at random, offsets in position, ellipticity, position angle and cut radius for each galaxy within the cluster with standard deviations of $\sigma_{\rm p} = 0.1\arcsec$, $\sigma_{\rm e}=20\%$, $\sigma_{\rm a}=20^\circ$ and $\sigma_{\rm cut}=10\%$. These values are physically motivated and based on the results of \cite{bary_dm_allign} and the offset observed in A3827 \citep{A3827_massey}.
We simulate the positions of the multiple images 20 times and find the mean positional offset of each image and the expected RMS given the assumption that light traces mass.
We also test the hypothesis that the images that are apparently closer to lenses on the sky will be the most sensitive to a misaligned dark matter halo.
Figure \ref{fig:all_offsets} gives the mean and one sigma standard error in the position offset for each multiple image as a function of the distance the image is from the closest lens. The solid black line shows the mean RMS over all realisations. The bars in each case give the approximate contribution from each effect to the total offset.
We find that assuming that light traces mass results in an estimated $\sim0.5\arcsec$ RMS error in the position of multiple images. 
The majority of the offset is contributed by a shift in the cut radius and angle position, however these appear not to depend on distance from the nearest lens.
We do find though that the position and ellipticity have a slight tendency to have a larger effect as the lens is closer to the image.
Given the accuracy of current HFF, the $0.5\arcsec$ here accounts for the majority of this error, and any future survey attempting to go below this limit will not be able to assume that mass traces light.

\subsection{Prospects of detecting misaligned halos in the Hubble Frontier Fields}
 Models in the Hubble Frontier Fields currently report a RMS error of $0.6\arcsec$ for a single cluster \citep{MACSJ0416_HFF}, of which most could be accounted for by assuming that light traces mass. 
The separation observed in A3827 was of order $\sim2$kpc, which is $\approx0.4\arcsec$ at a redshift of $z\sim0.397$. We found that should all galaxies have their dark matter component coherently offset by this amount we would observe a mean shift of $\sim0.05-0.1\arcsec$, which is currently beyond the accuracy of strong lensing models. However, given that this is based on one cluster and individual images very close to a lens can be shifted significantly, it is highly likely that future Frontier Fields should observe a configuration of multiple images that are sensitive to the offset between dark matter and baryons in particular lenses.
Apart from this, cosmological simulations predict that the position angle of halos can be misaligned by large amounts and dark matter halos can be rounder than their baryonic counterparts by factors of two. The predicted shift in multiple images for these two properties means that it is it is possible to make a first detection of misalignment in the Frontier Fields. Having said this, the movement of multiple images with respect to each offset is highly degenerate, and hence in order to characterise the statistical properties of the misalignments it will require a high density of multiple images around the lenses. Such a discovery would have far-reaching implications for galaxy formation models and also intrinsic alignments that have become a vital systematic in the measurement of cosmological gravitational lensing (see \cite{intrinsic_allign} for review).

\fig       
\includegraphics[width=0.5\textwidth]{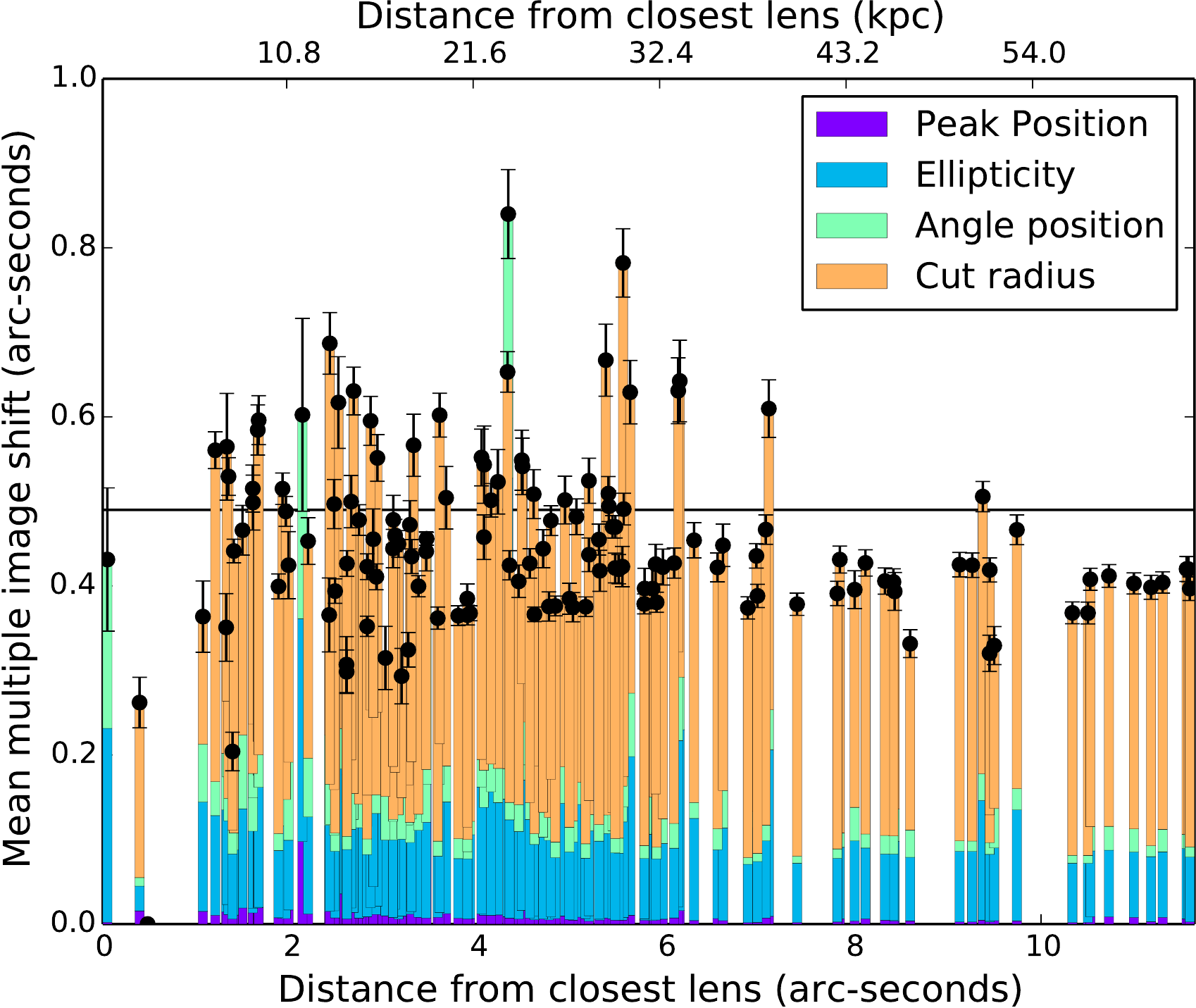}	
\caption{ \label{fig:all_offsets} We include variations in all parameters associated with assuming that mass traces light and randomly offset the dark matter halos in four ways in order to measure the expected RMS in the multiple image position. Using $\sigma_{\rm p} = 0.1\arcsec$, $\sigma_{\rm e}=20\%$, $\sigma_{\rm a}=20^\circ$ and $\sigma_{\rm cut}=10\%$ all with means of zero, and iterated over 20 realisations. We measure the mean and variance in the position of each multiple image and the {\it expected RMS} as a function of distance from the closest lens in angular and proper distances. We show the mean RMS over all the images with the solid black line. We also determine the approximate contribution for each parameter to the mean image shift}
\efig

\section{Conclusion}
The assumption that light traces mass is often used when modeling the distribution of mass in galaxy clusters.
Strong gravitational lensing models, which assume this report root mean square errors between the predicted position of multiple images and actual positions of $\sim 0.6\arcsec$.
In this study we test the validity of this assumption by altering the dark matter halo of galaxies in a model based on the Hubble Frontier Field cluster MACSJ0416 whilst keeping the large-scale cluster component and stellar component fixed. 
Assuming that light traces mass often requires the peak dark matter halo and baryonic halo to be exactly coincident, the ellipticity and the position angle of the halo to be aligned and the cut radius to have some kind of relation with the luminosity. 
In this study we test all four assumptions individually.
We find whether we commonly shift dark matter halos or individually randomly offset them, each individual case produces a mean offset of $\sim0.2\arcsec$ in the position of multiple images.
We find that although the mean offset for each image is of order $\sim0.2 \arcsec$, some images can be very sensitive to perturbations in the galactic dark matter halo with some misalignments resulting in a multiple image shift of $\sim1\arcsec$.

Following this we study how individual massive lenses in close proximity to images can significantly perturb the image position finding that some individual galaxy misalignments can induce image shifts of $>1\arcsec$.
Additional, we also find that the peak position, position angle and ellipticity offset result in a tangential movement of images whereas a change in the cut radius results in a radial movement, meaning that future studies should be able to discern between misalignments in the halo and departures from mass - luminosity relations.

We finally combine all four effects and find that the mass to light assumption can result in $\sim0.5\arcsec$ root mean square error in the position of multiple images, almost the entire error budget of the Frontier Field lensing models.

Given that misalignments are of physical interest to galaxy formation models and as a systematic error for measurements of cosmic shear, we find that given the current sensitivity and depth of the HFF, it should be possible to detect and characterise misalignments between dark matter and baryonic halos in the Hubble Frontier Field galaxy clusters.
We also find that given an expected offset of $\sim2$kpc between dark matter and baryons, the smoking gun for self-interacting dark matter, it maybe possible to detect some offset in the Hubble Frontier Fields, given the correct multiple image configuration.

\section*{Acknowledgements}
The authors would like to thank David Martin and Matthew Nichols for valuable advice in constructing the MNH figure.
DH is supported by the Swiss National Science Foundation (SNSF).
JPK acknowledges support from the ERC advanced grant LIDA and from CNRS.
MJ  acknowledge support by the Science and Technology Facilities Council [grant number ST/L00075X/1 \& ST/F001166/1]

\bibliographystyle{mn2e}
\bibliography{bibliography}

\bsp

\label{lastpage}

\end{document}